\documentclass[12pt]{article}
\usepackage{amsfonts}
\usepackage{amssymb}%,letterspace}
\usepackage{graphics,psboxit,amsmath}
\usepackage{subfigure}
\usepackage{graphicx}

\def\hybrid{\topmargin -30pt    \oddsidemargin 0pt %%%%%%%%%%%%%% Archive-30pt
        \headheight 0pt \headsep 0pt
        \textwidth 6.25in       % A4 paper
        \textheight 9.5in       % A4 paper
        \marginparwidth .875in
        \parskip 5pt plus 1pt   \jot = 1.5ex}

\hybrid

\def\baselinestretch{1.2}

\catcode`\@=11

\def\marginnote#1{}
\newcount\hour
\newcount\minute
\newtoks\amorpm
\hour=\time\divide\hour by60
\minute=\time{\multiply\hour by60 \global\advance\minute by-\hour}
\edef\standardtime{{\ifnum\hour<12 \global\amorpm={am}%
        \else\global\amorpm={pm}\advance\hour by-12 \fi
        \ifnum\hour=0 \hour=12 \fi
        \number\hour:\ifnum\minute<10 0\fi\number\minute\the\amorpm}}
\edef\militarytime{\number\hour:\ifnum\minute<10 0\fi\number\minute}
\def\draftlabel#1{{\@bsphack\if@filesw {\let\thepage\relax
   \xdef\@gtempa{\write\@auxout{\string
      \newlabel{#1}{{\@currentlabel}{\thepage}}}}}\@gtempa
   \if@nobreak \ifvmode\nobreak\fi\fi\fi\@esphack}
        \gdef\@eqnlabel{#1}}
\def\@eqnlabel{}
\def\@vacuum{}
\def\draftmarginnote#1{\marginpar{\raggedright\scriptsize\tt#1}}

\def\draft{
        \def\@oddfoot{\sl preliminary draft \hfil
        \rm\thepage\hfil\sl\today\quad\militarytime}
        \let\@evenfoot\@oddfoot \overfullrule 3pt
        \let\label=\draftlabel
        \let\marginnote=\draftmarginnote
   \def\@eqnnum{(\theequation)\rlap{\kern\marginparsep\tt\@eqnlabel}%
\global\let\@eqnlabel\@vacuum}  }

\def\preprint{\twocolumn\sloppy\flushbottom\parindent 2em
        \leftmargini 2em\leftmarginv .5em\leftmarginvi .5em
        \oddsidemargin -.5in    \evensidemargin -.5in
        \columnsep .4in \footheight 0pt
        \textwidth 10.in        \topmargin  -.4in
        \headheight 12pt \topskip .4in
        \textheight 6.9in \footskip 0pt
        \def\@oddhead{\thepage\hfil\addtocounter{page}{1}\thepage}
        \let\@evenhead\@oddhead \def\@oddfoot{} \def\@evenfoot{} }

\def\numberbysection{\@addtoreset{equation}{section}
        \def\theequation{\thesection.\arabic{equation}}}

\def\underline#1{\relax\ifmmode\@@underline#1\else
        $\@@underline{\hbox{#1}}$\relax\fi}

\def\titlepage{\@restonecolfalse\if@twocolumn\@restonecoltrue\onecolumn
     \else \newpage \fi \thispagestyle{empty}\c@page\z@
        \def\thefootnote{\fnsymbol{footnote}} }

\def\endtitlepage{\if@restonecol\twocolumn \else \newpage \fi
        \def\thefootnote{\arabic{footnote}}
        \setcounter{footnote}{0}}  %\c@footnote\z@ }

\catcode`@=12
\relax

\def\figcap{\section*{Figure Captions\markboth
        {FIGURECAPTIONS}{FIGURECAPTIONS}}\list
        {Figure \arabic{enumi}:\hfill}{\settowidth\labelwidth{Figure
999:}
        \leftmargin\labelwidth
        \advance\leftmargin\labelsep\usecounter{enumi}}}
 \relax
\def\tablecap{\section*{Table Captions\markboth
        {TABLECAPTIONS}{TABLECAPTIONS}}\list
        {Table \arabic{enumi}:\hfill}{\settowidth\labelwidth{Table
999:}
        \leftmargin\labelwidth
        \advance\leftmargin\labelsep\usecounter{enumi}}}
 \relax
\def\reflist{\section*{References\markboth
        {REFLIST}{REFLIST}}\list
        {[\arabic{enumi}]\hfill}{\settowidth\labelwidth{[999]}
        \leftmargin\labelwidth
        \advance\leftmargin\labelsep\usecounter{enumi}}}
 \relax
\makeatletter
\newcounter{pubctr}
\def\publist{\@ifnextchar[{\@publist}{\@@publist}}
\def\@publist[#1]{\list
        {[\arabic{pubctr}]\hfill}{\settowidth\labelwidth{[999]}
        \leftmargin\labelwidth
        \advance\leftmargin\labelsep
        \@nmbrlisttrue\def\@listctr{pubctr}
        \setcounter{pubctr}{#1}\addtocounter{pubctr}{-1}}}
\def\@@publist{\list
        {[\arabic{pubctr}]\hfill}{\settowidth\labelwidth{[999]}
        \leftmargin\labelwidth
        \advance\leftmargin\labelsep
        \@nmbrlisttrue\def\@listctr{pubctr}}}
 \relax
\makeatother
\newskip\humongous \humongous=0pt plus 1000pt minus 1000pt

\newif\ifdtup

\relax

\def\be{\begin{equation}}
\def\ee{\end{equation}}
\def\ba{\begin{eqnarray}}
\def\ea{\end{eqnarray}}

\def\g{\gamma}
\def\G{\Gamma}
\def\d{\delta}
\def\D{\Delta}

\def\P{\Pi}

\def\th{\theta}
\def\Th{\Theta}
\def\m{\mu}

\def\Om{\Omega}
\def\l{\lambda}
\def\L{\Lambda}
\def\s{\sigma}
\def\S{\Sigma}

\def\cN{{\cal N}}

\def\elK{{\bf K}}

\def\no{\noindent}

\def\qq{\qquad}

\def\IR{\relax{\rm I\kern-.18em R}}

\def \ha {{1\over 2}}

\def \ov {\over}

\def\const{{\rm const.}}

\def\II{\relax{\rm 1\kern-.35em1}}
\def\IR{\relax{\rm I\kern-.18em R}}
\def\inv{^{\raise.15ex\hbox{${\scriptscriptstyle -}$}\kern-.05em 1}}

\def\tL{{\tilde L}}

\begin{document}
%\draft

%\renewcommand{\theequation}{\arabic{equation}}
%\renewcommand{\theequation}{\thesection.\arabic{equation}}

\renewcommand{\theequation}{\thesection.\arabic{equation}}
\csname @addtoreset\endcsname{equation}{section}

\newcommand{\beq}{\begin{equation}}
\newcommand{\eeq}[1]{\label{#1}\end{equation}}
\newcommand{\ber}{\begin{eqnarray}}
\newcommand{\eer}[1]{\label{#1}\end{eqnarray}}
\newcommand{\eqn}[1]{(\ref{#1})}
\begin{titlepage}
\begin{center}

%\hfill CERN-PH-TH/2005-192\\
%\vskip -.1 cm
\hfill hep--th/0606190\\

\vskip .5in

{\Large \bf Supergravity and the jet quenching parameter\break
in the presence of R-charge densities}

\vskip 0.5in

%\begin{comment}
{\bf Spyros D. Avramis$^{1,2}$} \phantom{x} and\phantom{x}  {\bf Konstadinos Sfetsos}$^2$
\vskip 0.1in

${}^1\!$
Department of Physics, National Technical University of Athens,\\
15773, Athens, Greece\\

\vskip .1in

${}^2\!$
Department of Engineering Sciences, University of Patras,\\
26110 Patras, Greece\\

\vskip .2in

{\footnotesize {\tt avramis@mail.cern.ch}, \ \ {\tt sfetsos@upatras.gr}}\\

\end{center}

\vskip .4in

\centerline{\bf Abstract}

\no
Following a recent proposal,
we employ the AdS/CFT correspondence to compute the jet quenching
parameter for $\cN=4$ Yang--Mills theory at nonzero R-charge
densities. Using as dual supergravity backgrounds non-extremal
rotating branes, we find that the presence of the R-charges
generically enhances the jet quenching phenomenon. However, at
fixed temperature, this enhancement might or might not be a
monotonically increasing function of the R-charge density and
depends on the number of independent angular momenta describing
the solution. We perform our analysis for the canonical as well as
for the grand canonical ensemble which give qualitatively similar
results.

\no

%\vfill
%\vskip .5cm
%\noindent

\end{titlepage}
\vfill
\eject

\def\baselinestretch{1.2}
\baselineskip 10 pt
\noindent

%%%%%%%%%%%%%%%%%%%%%%%%A generalisation of target space%%%%%%%%%%%%%
\def\tT{{\tilde T}}
\def\tg{{\tilde g}}
\def\tL{{\tilde L}}

%%%%%%%%%%%%%%%%%%%%%%%%%%%%%%%%%%%%%%%%%%%%%%%%%%%%%%%%%%%%%%%%%%%%%%
%%%%%%%%%%%%%%%%%%%%%%%%%%%%%%%%%%%%%%%%%%%%%%%%%%%%%%%%%%%%%%%%%%%%%%

%\tableofcontents

\def\baselinestretch{1.2}
\baselineskip 20 pt
\no

\section{Introduction and summary}

There exists strong evidence from RHIC data that the hot, dense
QCD plasma produced at the temperature range $T_{\rm c} < T < 4
T_{\rm c}$, where $T_{\rm c}$ is the crossover temperature,
remains strongly coupled ($g^2 N \simeq 10$) despite the partial
deconfinement of color charges (see e.g. \cite{qgpfluid} for a
review). As such, its behavior is similar to that of a nearly
perfect fluid and its macroscopic properties admit an effective
description in terms of hydrodynamics \cite{qgpfluid2}. However,
the hydrodynamic parameters characterizing the fluid are
notoriously hard to compute using conventional approaches and,
being of dynamical nature, they are not amenable to lattice
calculations.

\no On the other hand, the fact that the theory is strongly
coupled in the above temperature range suggests the use of
nonperturbative gauge/gravity dualities for such computations.
Unfortunately, such dualities do not apply to real QCD but rather
to its supersymmetric generalizations, the prototype example being
the AdS/CFT correspondence \cite{adscft} relating Type IIB string
theory on $AdS_5 \times S^5$ to $\cN = 4$ SYM. Nevertheless, there
is some hope that, in a strongly-coupled yet non-confining regime,
some basic aspects of QCD dynamics may be captured by a
supersymmetric theory possessing a gravity dual. This line of
approach was employed in a series of works \cite{starinets} for
the calculation of transport coefficients which yielded the
remarkable result that the ratio of shear viscosity to entropy
density attains a universal value, close to the observed one, in any
theory with a gravity dual \cite{svuniversality} (see also
\cite{viscosity-rcharge}). Such encouraging results provide enough
motivation for trying to apply, with the proper caution,
gauge/gravity dualities in the study of phenomena related to the
QCD plasma.

\no An interesting phenomenon in the above class is jet quenching,
the energy loss of high--$p_T$ partons produced in heavy-ion
collisions as they interact with the plasma before they fragment
into hadrons \cite{jqphen,kovner}. The ``transport coefficient''
characterizing the phenomenon is the \emph{jet quenching
parameter} $\hat{q}$, usually defined perturbatively as the
average loss in four-momentum squared per mean free path. By
analogy to high-energy scattering studied in the framework of
eikonal approximations (see e.g. \cite{kovner}), the problem of
jet quenching can be formulated \cite{wiedemann} in terms of
Wilson loops in the adjoint representation. In such a context,
$\hat{q}$ can be calculated according to the relation
\be
\label{aa} \langle W^A (C) \rangle\ = \exp \left( - {1 \ov 4}
\hat{q}L^- L^2 \right)\ ,
\ee
where $W^A (C)$ is an adjoint Wilson loop on a rectangular contour
$C$ with one lightlike and one spacelike side of lengths $L^-$ and
$L$ respectively, with $L \ll L^-$.

\no The fact that Wilson loops of this type lend themselves to
strong-coupling calculations via gauge/gravity dualities
\cite{wilsonloop,wilsonloop2,BS} motivated Liu, Rajagopal and
Wiedemann \cite{liu} to postulate that the above relation may be
taken as a nonperturbative operational definition of $\hat{q}$,
and to calculate its value in $\cN=4$ SYM according to AdS/CFT. At
large $N$, one may take $\langle W^A (C) \rangle \simeq \langle
W^F (C) \rangle^2$, where $W^F(C)$ is a fundamental Wilson loop,
and use the AdS/CFT relation $\langle W^F (C) \rangle = \exp
\left( {\rm i} S[C] \right)$, where $S[C]$ is the Nambu-Goto
action for a string propagating in the dual gravity background and
whose endpoints trace the contour $C$. Noting that the exponent in
\eqn{aa} is real, the action $S[C]$ must be {\it imaginary}, i.e.
the string configuration of interest must be {\it spacelike}.
Writing $S[C] = {\rm i} \tilde{S}[C]$, where $\tilde{S}[C]$ is
real, we find that $\hat{q}$ is determined by
\be
\label{1-1} \frac{1}{8} \hat{q} L^- L^2 = \tilde{S}[C]\ ,
\ee
in the limit of small $L$. The meaning of the above relation is
that one should seek a solution of the string equations of motion
with the string endpoints lying on the lightlike loop $C$ and,
provided that the leading term in the small-$L$ expansion of its
action is proportional to $L^2$, read off the coefficient
$\hat{q}$ from \eqn{1-1}.

\no Two other observables related to the phenomenon of energy loss
in plasmas is the drag force exerted on a heavy quark as it
travels through the plasma and the heavy-quark diffusion
coefficient; in the context of AdS/CFT, these have been calculated
in \cite{drag1,drag2} and \cite{diffusion} respectively while the
first approach was further explored in \cite{mich,herzog,caceres}.

\no The validity of the approach described above for the
computation of $\hat{q}$ and its relation to the ``drag force''
approach to the study of parton energy loss has been a subject of
much debate in the recent literature \cite{guijosa,lrwlong,aev}.
To begin, the two approaches are conceptually completely different
as the proposal of \cite{liu} corresponds to a limiting case of a
quark-antiquark string solution \cite{lrw-wind} whose action is
required to be imaginary, while that of \cite{drag1,drag2} refers
to a single-quark string solution with the whole computation based
on the requirement that the action be real. In this respect, it is
now quite clear \cite{drag1,lrwlong} that the two types of
approach describe different physics, with the approach of
\cite{liu} presumably best suited to light quarks and that of
\cite{drag1,drag2} best suited to heavy quarks. Nevertheless, it
is still argued \cite{aev} that the implicit limiting procedure
employed in \cite{liu} is not valid and that the spacelike string
used there does not dominate the path integral. Although such
points will not be further discussed here, we must stress that the
computations presented in this paper are valid provided that the
proposal of \cite{liu} is on solid grounds.

\no Be that as it may, the value of $\hat{q}$ computed in
\cite{liu} is of the correct order of magnitude, and the obvious
question is how it is modified in more generalized settings, as
done for example in \cite{buchel} for certain non-conformal cases.
In this paper we compute the jet quenching parameter, by means of
the method outlined above, for the case of $\cN=4$ SYM in the
presence of nonzero R-charges. In section 2, we review the dual
supergravity backgrounds, corresponding to non-extremal rotating
D3-branes, and we concentrate on two cases with one or two equal
nonzero angular momenta, for which we state the criteria for
thermodynamic stability and the relations between the gauge-theory
and supergravity parameters. In section 3, we compute exactly the
jet quenching parameter $\hat{q}$ as a function of the
thermodynamic parameters of the gauge theory, in both the
canonical and grand canonical ensembles. Our main conclusion is
that turning on nonzero R-charges generically enhances the jet
quenching phenomenon, in a manner dependent on the number of
non-vanishing equal angular momenta.
\section{Non-extremal rotating branes}

According to the AdS/CFT correspondence, $\cN=4$ super Yang-Mills
theory with $SU(N)$ gauge group is dual to a stack of $N$ extremal
D3-branes. Introducing finite temperature corresponds to replacing
the extremal branes by non-extremal ones \cite{adscftfiniteT},
while introducing nonzero R-charges corresponds to generalizing
the branes to rotating ones. This class of metrics has been found
in full generality in \cite{rotatingbranesmore} using previous
results from \cite{cvetic}. They are characterized by the
non-extremality parameter $\m$ plus the rotation parameters $a_i$,
$i=1,2,3$, which correspond to the three generators of the Cartan
subalgebra of $SO(6)$ and are related to three chemical potentials
(or R-charges) in the gauge theory. The most general non-extremal
rotating D3-brane solution in the field-theory limit is given by
\cite{rs}
\ba
&& ds^2 = H^{-1/2} \left[-\left(1-{\m^4\ov r^4 \D}\right) dt^2 + dx^2 + dy^2 + dz^2\right]
+ H^{1/2}{r^6 \D \ov f}\ dr^2
\nonumber\\
&& \, + \:\: H^{1/2}
\Bigg[r^2 \Delta_1 d\theta^2 + r^2 \Delta_2 \cos^2\theta d\psi^2 +
2 (a_2^2-a_3^2)\cos\theta\sin\theta\cos\psi\sin\psi d\theta d\psi
\label{dsiib}\\
&& \, + \:\: (r^2+a_1^2) \sin^2\theta d\phi_1^2 +
(r^2+a_2^2) \cos^2\theta \sin^2\psi d\phi_2^2 +
(r^2+a_3^2) \cos^2\theta\cos^2\psi d\phi_3^2
\nonumber\\
&& -\ 2 {\m^2\ov R^2} \ dt \ (a_1 \sin^2 \th\ d \phi_1 + a_2 \cos^2 \th \sin^2\psi\ d \phi_2
+ a_3 \cos^2 \th \cos^2\psi\ d \phi_3 )
\Bigg] \ ,
\nonumber
\ea
where the diverse functions are defined as
\ba
H & = & {R^4\ov r^4 \D}\ ,
\nonumber\\
f & = & (r^2+a_1^2)(r^2+a_2^2)(r^2+a_3^2)- \m^4 r^2\ ,
\nonumber\\
\Delta &=& 1 +{a_1^2\over r^2} \cos^2\theta +{a_2^2\over r^2}
(\sin^2\theta\sin^2\psi +\cos^2\psi )
+ {a_3^2\over r^2}(\sin^2\theta\cos^2\psi +\sin^2\psi )
\nonumber \\
&+& {a_2^2 a_3^2\over r^4}\sin^2\theta +{a_1^2 a_3^2\over r^4}
\cos^2\theta\sin^2\psi +{a_1^2 a_2^2 \over r^4}\cos^2\theta\cos^2\psi\ ,
\label{d12}\\
\Delta_1
 &=& 1+{a_1^2\ov r^2}\cos^2\theta +
{a_2^2\ov r^2}\sin^2\theta\sin^2\psi +
{a_3^2\ov r^2}\sin^2\theta\cos^2\psi\ ,
\nonumber\\
\Delta_2 &=& 1+{a_2^2\ov r^2}\cos^2\psi +{a_3^2\ov r^2}\sin^2\psi\ .
\nonumber
\ea
This metric is supported by a self-dual 5-form $F_5 = dC_4+\star dC_4$ with potential
\ba
&&\!\!\!\!\!\!\!\! C_4  =  C_1 \wedge dx \wedge dy \wedge dz\ ,
\nonumber\\
&&\!\!\!\!\!\!\!\! C_1 = - H^{-1} dt
+  {\m^2\ov R^2} \ (a_1 \sin^2 \th\ d \phi_1 + a_2 \cos^2 \th \sin^2\psi\ d \phi_2
+ a_3 \cos^2 \th \cos^2\psi\ d \phi_3 )\ .
\label{dsj1}
\ea
The location $r_H$ of the horizon is defined as the largest root of the cubic,
in $r^2$, equation
\be
f  =  (r^2+a_1^2)(r^2+a_2^2)(r^2+a_3^2)- \m^4 r^2 = 0\ .
\label{hoot}
\ee
The thermodynamic properties of this metric were worked out
in \cite{rotatingbranesthermo1}-\cite{rotatingbranesthermo3}, \cite{rs}.
The Hawking temperature, entropy, energy above extremality,
angular velocities and angular momenta read
\ba
&&T = {r_H\ov 2 \pi \m^2 R^2} \left( 2 r_H^2 + a_1^2 + a_2^2 + a_3^2
- {a_1^2 a_2^2 a_3^2 \ov r_H^4} \right)\ ,
\nonumber\\
&&S =  {N^2\m^2 r_H\ov 2\pi R^6}\ ,\qq E={3N^2\m^4\ov 8\pi^2 R^8}\ ,
\label{genneHa}
\\
&& \Om_i = {a_i \ov a_i^2+r_H^2} {\m^2\ov R^2} \ ,\qq
J_i = {N^2 \m^2 a_i\ov 4\pi^2 R^6}\ ,\qq i=1,2,3\ ,
\nonumber
\ea
where the various extensive quantities are understood as the
respective densities. We must here stress that, in extracting
information about the gauge theory, one must trade the
supergravity parameters ($\m$,$a_i$) for the parameters
($T$,$J_i$) or ($T$,$\Omega_i$) which have a direct gauge-theory
interpretation with $J_i$ and $\Omega_i$ playing the role of
R-charge densities and chemical potentials respectively; choosing
($T$,$J_i$) corresponds to the Canonical Ensemble (CE) while
choosing ($T$,$\Omega_i$) corresponds to the Grand Canonical
Ensemble (GCE). The gauge-theory and supergravity parameters are
generally not in one-to-one correspondence and their physical
range is determined by thermodynamic stability.

\no In the rest of this paper, we will consider in detail two
cases, parametrized by the non-extremality parameter $\m$ and one
or two angular momentum parameters which are set to a common value
$r_0$. For these spinning branes, the corresponding angular
momenta and velocities are equal and the criteria for
thermodynamic stability are easily stated, translating into the
following upper bound for the common angular momentum
\cite{rotatingbranesthermo1}-\cite{rotatingbranesthermo3}
\be
J \leqslant \sqrt{x^{(m)}_c} {S\ov 2\pi}\ ,
\label{jh1}
\ee
where the coefficients $x^{(m)}_c$ depend on the number $m$ of equal angular momenta,
as well as on whether we utilize the CE or GCE. In particular we recall
from table 4.1 of \cite{rotatingbranesthermo3} that
\ba
{\rm CE} :&& \qq x^{(1)}_c={5+\sqrt{33}\ov 12}\ ,\qq x_c^{(2)}=\infty\ ,
\nonumber\\
{\rm GCE} :&& \qq x^{(1)}_c=2\ ,\qq x_c^{(2)}=1\ .
\label{sdk1}
\ea
We note that increasing the number of equal angular momenta
stabilizes the D3-branes, as indicated here by the disappearance
of the bound for the case of two equal angular momenta in the CE.
In addition, from the same reference we recall that the parametric
space spanned by $(\m,r_0)$ is such that
\ba
{\rm CE} && : \qq j={64 J^2\ov \pi^6 N^4 T^6}\leqslant j^{(m)}_c\ ;
\qq j^{(1)}_c \simeq 14.12\ ,\quad j^{(2)}_c=\infty\ , \nonumber\\
{\rm GCE} && : \qq {\Om\ov T} \leqslant  \pi a^{(m)}_c \ ;
\qq a^{(1)}_c={1\ov \sqrt{2}}\ ,\quad a^{(2)}_c=1\ .
\label{hf1}
\ea

\subsection{Two equal nonzero angular momenta}

We first examine the case of two equal nonzero rotation
parameters, which we can take as $a_2=a_3=r_0$ with $a_1=0$. The
metric \eqn{d12} simplifies to
\ba
&& ds^2 = H^{-1/2} \left[-\left(1-{\m^4 H \ov R^4}\right) dt^2 + dx^2 + dy^2 + dz^2\right]
+ H^{1/2} {r^4(r^2-r_0^2 \cos^2\th)\ov (r^4-\m^4)(r^2-r_0^2)}\ dr^2
\nonumber\\
&& + H^{1/2}\Big[(r^2-r_0^2\cos^2\th )d\th^2 + r^2 \cos^2\th d\Om_3^2
+ (r^2-r_0^2)\sin^2\th d \phi_1^2
\label{sphe}
\\
&& - \ 2 {\m^2 r_0\ov R^2} \ dt \cos^2\th (\sin^2\psi d \phi_2 + \cos^2\psi d \phi_3)\Big]\ ,
\nonumber
\ea
where
\be
H={R^4\ov r^2 (r^2-r_0^2\cos^2 \th)} \
\ee
and the line element of the three-sphere is
\be
d\Om^2_3  =d\psi^2 + \sin^2 \psi \ d \phi_2^2 + \cos^2 \psi \ d \phi_3^2 \ .
\label{3spp}
\ee
We have also shifted the radial coordinate as $r^2\to r^2 -r_0^2$.
Taking this into account, we compute from \eqn{hoot} the location
of the horizon at
\be
\label{2-4}
r_H = \m\ ,
\ee
while the various thermodynamic quantities read
\ba
&&T = {\sqrt{\m^2 - r_0^2} \ov \pi R^2}\ ,\qq S = {N^2\m^2 \sqrt{\m^2-r_0^2}\ov 2\pi R^6}\ ,
\nonumber\\
&&\Om = {r_0\ov R^2}\ ,\qq J = {r_0 \m^2 N^2 \ov 4 \pi^2 R^6}\ .
\label{dhj1}
\ea
The reality condition $\m\geqslant r_0$ should be imposed. In what
follows we will need to solve Eqs. \eqn{dhj1} for $\m$ and $r_0$
in terms of the pairs $(T,J)$ or ($T,\Om)$, relevant for the gauge
theory, in the CE and GCE, respectively. Luckily, in this case
these equations can be solved exactly for both ensembles.

\no $\bullet$
For the CE case the result is most conveniently
expressed in terms of the dimensionless parameter
\ba
\label{3-18}
\xi & = &  {6 \sqrt{3} \ov \pi N^2}{J \ov T^3}\ ,
\nonumber\\
& = & {3\sqrt{3}\ov 2} { \l \ov (1-\l^2)^{3/2}}\ ,\qq \l = {r_0\ov \m}\
\ea
and a function $F$ introduced so that the parameters
$\mu$ and $r_0$ are given by
\be
\label{3-20}
\m^2 = {\pi^2 R^4 T^2} \left( 1 + F^2 \right)\ ,\qq
r_0 = \pi R^2 T F\  .
\ee
Then the equation for $T$ in \eqn{dhj1} is identically satisfied
while a substitution into the equation for $J$ yields the algebraic equation
\be
F(F^2+1)= {2\ov 3 \sqrt{3}}\ \xi\ .
\ee
Its real solution is given by
\be
\label{3-19}
F(\xi) ={(\xi + \sqrt{1+\xi^2})^{1/3} - (\xi + \sqrt{1+\xi^2})^{-1/3}\ov \sqrt{3}}\ .
\ee
Note that, since $0\leqslant \l \leqslant 1$, the parameter $0\leqslant \xi< \infty$
monotonically.
Also, for small $\xi$, it admits the expansion
\be
F(\xi)= {1\ov \sqrt{3}}\left({2\ov 3}\xi - {8\ov 81} \xi^3 + {32\ov 729}\xi^5\right)
+ {\cal O}(\xi^7)\ ,
\ee
while for large $\xi$ it behaves as $F(\xi)\simeq 2^{1/3} 3^{-1/2} \xi^{1/3}$.
In this case, as we see from the stability conditions \eqn{jh1}-\eqn{hf1},
no restrictions are placed on the various parameters except the obvious one $\m\geqslant r_0$, mentioned above.

\no
$\bullet$
For the case of the GCE, it is convenient to employ the dimensionless parameter
\ba
\hat \xi ={\Om\ov \pi T} ={\l\ov \sqrt{1-\l^2}}\ ,\qq \l={r_0\ov \m}\ .
\label{fhi2}
\ea
Introducing the function $F$ as in \eqn{3-20}, we simply find
$F(\hat \xi)= \hat \xi$.
In this case the stability conditions \eqn{jh1}-\eqn{hf1}
require that
\be
\hat \xi \leqslant \hat \xi_{\rm max} = 1 \quad {\rm and}\quad   \l\leqslant {1\ov \sqrt{2}}\ ,
\ee
which are in fact equivalent.
Note that, were it not for the thermodynamic stability considerations, there would not
be any a priori reason for restricting the range of $\hat \xi$ as above.

\no For $\m=0$ the background metric \eqn{sphe} corresponds to a
uniform distribution of D3-branes on a three-dimensional spherical
shell with radius $r_0$ on $\mathbb{R}^4$ and supersymmetry is
restored. In the limit $\m\to 0$, the thermodynamic description
becomes meaningless as the temperature and entropy attain
imaginary values.

\subsection{One nonzero angular momentum}

We next examine the case of only one nonzero rotation parameter,
which we can take as $a_1=r_0$ with $a_2=a_3=0$. The metric
\eqn{d12} simplifies to
\ba
&& ds^2 = H^{-1/2} \left[-\left(1-{\m^4 H \ov R^4}\right) dt^2 + dx^2 + dy^2 + dz^2\right]
+ H^{1/2} {r^2(r^2+r_0^2 \cos^2\th)\ov r^4+r_0^2 r^2 -\m^4}\ dr^2
\nonumber\\
&& + H^{1/2}\Big[(r^2+r_0^2\cos^2\th )d\th^2 + r^2 \cos^2\th d\Om_3^2
+ (r^2+r_0^2)\sin^2\th d \phi_1^2
\label{discc}
\\
&& - \ 2 {\m^2 r_0\ov R^2}  \sin^2\th dt d\phi_1\Big]\ ,
\nonumber
\ea
where
\be
H={R^4\ov r^2 (r^2+r_0^2\cos^2 \th)} \
\ee
and the line element of the three-sphere is as in \eqn{3spp} above.
Now the horizon is located at $r=r_H$ with
\be
\label{2-9}
r_H^2 = {1 \ov 2} \left( - r_0^2 + \sqrt{r_0^4 + 4 \m^4} \right)\ ,
\ee
while the various thermodynamic quantities read
\ba
%\label{tsoj}
&&T =  {r_H \sqrt{r_0^4 + 4 \m^4} \ov 2 \pi R^2 \m^2}\ ,\qq S = {N^2\m^2 r_H\ov 2\pi R^6}\ ,
\nonumber\\
&&\Om = {r_0r_H^2\ov R^2\m^2} \ ,\qq J = {r_0 \m^2 N^2 \ov 4 \pi^2 R^6}\ .
\label{jhsd}
\ea
Again, we need to solve Eqs. \eqn{jhsd} for $\m$ and $r_0$ in terms of $(T,J)$ or ($T,\Om)$.

\no
$\bullet$
For the case of the CE we cannot explicitly solve for the parameters
$\m$ and $r_0$ in terms of $T$ and $J$.
Nevertheless this can be done perturbatively as follows.
Introducing a dimensionless parameter $\xi$ given by
\ba
\label{xidef}
\xi & = & {2 \sqrt{2} \ov \pi N^2}{J\ov  T^3}\ ,
\nonumber\\
& = & {16\l \ov (\l^4+4)^{3/2}(\sqrt{\l^4+4}-\l^2)^{3/2}}\ , \qq \l={r_0\ov \m}\ .
\ea
and a function $F$ such that
\be
\m^4 = \pi^4 R^8 T^4  F^3 (2-F)\ ,\qq  r_0^2 = 2\pi^2 R^4 T^2  F (F-1)\ ,
%\qq a= \sqrt{2} \pi R^2 T\,
\label{kpr1}
\ee
the equation for $T$ in \eqn{jhsd} is identically satisfied. Substitution into the equation for
$J$ gives the algebraic equation
\be
F^4 (F-1)(F-2)+\xi^2 = 0 \ .
\label{hr3}
\ee
Examining \eqn{xidef} we note that the parameter $\xi$, regarded as a function of $\l$, goes to
 zero for small and large values of $\l$. In between them it reaches a maximum with
\be
\xi_{\rm max}=\left({2879+561\sqrt{33}\ov 3456}\right)^{1/2}\simeq 1.33\ ,
\qq  {\rm at}\quad \l_0 = \left(19+3 \sqrt{33}\ov 8\right)^{1/4}\simeq 1.46 \ .
\ee
This is consistent with the fact that \eqn{hr3} has no solutions
for large enough $\xi$. This maximum value is found by requiring
that, besides \eqn{hr3}, its first derivative w.r.t. $F$ vanishes
as well. These conditions give $F={15+\sqrt{33}\ov 12}\simeq 1.73$
at $\xi=\xi_{\rm max}$ as above, which is the maximum value for
which \eqn{hr3} has a solution. For $0\leqslant \xi < \xi_{\rm
max}$ the algebraic equation \eqn{hr3} has two solutions that can
be approximated for small values of $\xi$ by the perturbative
expansions
\ba
{\rm near}\ F(\xi) = 1 && :\qq F(\xi)= 1+\xi^2 -3 \xi^4+16 \xi^6
+ {\cal O}(\xi^8)\ ,\nonumber\\
{\rm near}\ F(\xi) = 2 && : \qq F(\xi)= 2-{1\ov 16} \xi^2
-{3\ov 256} \xi^4 - {29\ov 8192} \xi^6
+ {\cal O}(\xi^8)\ .
\ea
Note next that the stability condition \eqn{jh1} requires that
$\l\leqslant \l_0$ which implies that the second solutions
corresponding to the expansion near $F=2$ should be rejected. Also
note that the parameter in \eqn{hf1} is $j=8 \xi^2$ and its
maximum value evaluated using $\xi_{\rm max}$ is well approximated
by the result quoted in \eqn{hf1}.

\no $\bullet$ For the GCE case it is possible to explicitly solve
for the parameters $\m$ and $r_0$ in terms of $T$ and $\Om$.
Introducing the dimensionless parameter
\ba
 \hat \xi & = & {\Om\ov \sqrt{2}\pi T}
\nonumber\\
& = &
\l {(\sqrt{\l^4+4}-\l^2)^{1/2}\ov \sqrt{\l^4+4}}\ ,\qq \l={r_0\ov \m}\ ,
\label{ol1}
\ea
and the function $F$ defined as in \eqn{kpr1}, we find upon
substitution into the expression for $\Om$ the quadratic equation
\be
(F-1) (F-2) + \hat \xi^2= 0 \ ,
\ee
with solutions
\be
F=F_\pm(\hat \xi)={3\ov 2}\pm \sqrt{{1\ov 4}-\hat\xi^2}\ .
\ee
In this case the stability condition \eqn{jh1}
requires that
\be
\l\leqslant \left(4\ov 3\right)^{1/4}\simeq 1.075\ ,
\ee
which is in fact the value in which the maximum value of $\hat \xi$ in \eqn{ol1} is acquired and
is also in agreement with \eqn{hf1}.
Then it follows that
\be
0 \leqslant \hat \xi \leqslant   \hat \xi_{\rm max} = \ha\ ,
\ee
and that the solution $F_+$ is not stable.

\no For $\m=0$ the background metric \eqn{discc} corresponds to a
uniform distribution of D3-branes on a disc with radius $r_0$ and
supersymmetry is restored. In the limit $\m\to 0$ the
thermodynamic quantities in \eqn{jhsd} are finite, but, on the
other hand, this limit lies beyond the stability bounds.

\section{Computation of the jet quenching parameter}

In what follows, we will generalize the calculation of \cite{liu}
for the case of rotating non-extremal branes and we will calculate
the jet quenching parameter for the cases considered in Section 2.
This calculation corresponds to a special limiting case of the
Wilson loop calculation for a quark-antiquark pair moving with
velocity $v$ and located on a probe brane at $u = \L$, in which
one takes the lightlike limit $v \to 1$ before sending the probe
brane to infinity by taking $\L \to \infty$ \cite{lrw-wind}. For
details on the various salient points of the calculation, the
reader is referred to \cite{lrwlong,aev}.

\no In order to cover general situations we will consider a
general class of ten-dimensional metrics of the form
\be
ds^2 = G_{tt} dt^2 + G_{xx}dx^2 + G_{yy} dy^2 + G_{rr} dr^2 + G_{\th\th} d\th^2 + \cdots\ ,
\ee
where the ellipses denote other possible terms involving the
remaining five variables as well as mixed terms. We pass to the
light-cone coordinates $x^\pm = {1\ov \sqrt{2}}(t\pm x)$ and
consider a Wilson loop on a rectangular contour $C$ of sides $L^-$
and $L$ along $x^-$ and $y$ respectively; in the approximation
where (\ref{1-1}) is valid, we must have $L \ll L^-$. In the
supergravity approach, the expectation of the Wilson loop is given
by the extremum of the Nambu-Goto action for a string extending in
the internal space whose endpoints trace the contour $C$. To fix
reparametrization invariance, we can take $(\tau,\s) = (x^-,y)$.
Since $L^- \gg L$, we may assume translational invariance along
$x^-$, i.e. $x^\mu = x^\mu (y)$. The embedding of the string in
the background is described by the functions $u=u(y)$ and
$\th=\th(y)$ with all other coordinates set to
constants.\footnote{To conform with established notation in the
literature we will use $u$ instead of $r$ in the Wilson loop
computations.} We assume that this is consistent with the
equations of motion and indeed this is the case for our metrics.
In summary, the surface we are interested in is parametrized by
the embedding
\be
\label{3-1}
u=u(y)\ ,\qq \th=\th(y)\ ,\qq  x^+\ ,\dots  = \const\ ,
\ee
subject to the boundary condition $u(\pm L/2)\! \to\! \infty$. The
Nambu-Goto action for this configuration is then given by $S_1 =
{\rm i} \tilde{S}_1$ with\footnote{We may allow some of the angles
$\phi_i$ to depend on $y$ in a way consistent with the equations
of motion, without affecting much the resulting action below. In
these cases one effectively replaces $\th'^2$ by $\th'^2+\sin^2\th
\phi'^2$. This is only the upper cap of $S^2$ since
$0\leqslant\th\leqslant \pi/2$. More specifically,
$\phi=\phi_2=\phi_3$ and $\phi=\phi_1$ for the cases of two
angular momenta and one angular momentum, respectively.}
\be
\label{3-3} \tilde{S}_1 = { L^- \ov 2 \pi}  \int dy \sqrt{{f(u)} +
g(u) u^{\prime 2} + h(u) \th^{\prime 2} }\ ,
\ee
where the prime denotes a derivative with respect to $y$ and
\be
\label{3-4}
f(u) = {1\ov 2} (G_{xx}+G_{tt}) G_{yy}\ ,\quad
g(u) = {1\ov 2} (G_{xx}+G_{tt}) G_{rr}\ ,\quad
h(u) = {1\ov 2} (G_{xx}+G_{tt}) G_{\th\th}\ .
\ee
Some comments are in order here. First, although in all our
examples the metric components depend explicitly on $\theta$, in
the functions $f$, $g$ and $h$ defined in \eqn{3-4} all
$\theta$-dependence drops out. This fact, which does not hold in
static Wilson-loop calculations (see \cite{BS} for static Wilson
loops using rotating branes and their supersymmetric limits), is
the main reason that makes the rest of the calculation possible
(and straightforward) in the rotating-brane case. Second, in all
of our examples $f(u)$ attains a constant value which we will
henceforth denote by $f_0$.

\no The action (\ref{3-3}) is obviously independent of $y$ and
$\th$. This implies that the associated ``Hamiltonian'' and
``angular momentum'' are conserved, leading to the equations
\be
\label{3-5}
g u^{\prime 2} + h \th^{\prime 2} = f_0 {\g^2 }\ , \qq h \th^\prime = \d \ ,
\ee
where $\g^2$ and $\d$ are integration constants, the former chosen
to be positive semidefinite since $g(u)$ is positive for large
enough $u$.
%----------------------------------------------------------
%
%\no
%If we had chosen the more general embedding
%\ba
%\label{3-1}
%a_2=a_3=r_0 &:& u=u(y)\ ,\quad \th=\th(y)\ ,\quad \phi_2 = \phi_3 = \phi(y)\ , \qq x^+\ ,\dots  = \const\ ,\nonumber\\
%a_1=r_0 &:& u=u(y)\ ,\quad \th= {\pi \ov 2} - \th(y)\ ,\quad  \phi_1 = \phi(y)\ , \qq x^+\ ,\dots  = \const\ ,
%\ea
%the Nambu-Goto action would be
%\be
%\label{3-3}
%S_1 = { L^- \ov 2 \pi}  \int dy \sqrt{{f(u)} + g(u) u^{\prime 2}
%+ h(u) \left( \th^{\prime 2} + \sin^2\th \phi^{\prime 2} \right) }\ ,
%\ee
%\no
%The associated conservation laws would have the form
%\be
%\label{3-5}
%g u^{\prime 2} + h ( \th^{\prime 2} + \sin^2\th \phi^{\prime 2} ) \ = f_0 {\g^2 }\ , \qq h \sin^2\th \phi^{\prime} = \d \ ,
%\ee
%{\it NOTE: If I take $\psi$ and the remaining $\phi$'s to be also functions of $y$, the resulting expressions are not simple and they depend on whether I have one or two rotation parameters. I think we don't need to generalize the discussion that much.}
%----------------------------------------------------------
Since it will turn out that in our cases $h(u)\sim 1/u^2$, we will
take the constant $\d=0$ since otherwise we cannot reach the
boundary at $u=\infty$ and simultaneously preserve the reality of
the solution to the differential equation. Then this equation has
a solution where $u$ starts from $u(-L/2)=\infty$, decreases until
it reaches a turning point, and then increases until it returns
back to $u(L/2)=\infty$. The turning point corresponds to the
largest zero of $g^{-1}(u)$, denoted by $u_{\min}$, and, by
symmetry, must occur at $y=0$. Then, Eq. (\ref{3-5}) can be
integrated with the result
\be
\label{3-8}
L = {2\ov \sqrt{f_0} \ \g} \int_{u_{\min}}^\infty du \sqrt{g(u)}
= {2 \ov \sqrt{f_0} \ \g} I[g]\ .
\ee
where $I[g]$ denotes the explicit expression of the integral as a
functional of $g(u)$ that depends on the metric components as in
\eqn{3-4}. Meanwhile, given Eq. (\ref{3-5}), Eq. (\ref{3-3}) gives
for the action
\be
\label{3-9} \tilde{S}_1 = {\sqrt{f_0} L^-  L \ov 2  \pi } \sqrt{1
+ \g^2} \ ,
\ee
which is obviously real, implying that the solutions are spacelike
as mentioned in the introduction. From \eqn{3-9}, we must subtract
the ``self-energy'' contribution arising from the disconnected
worldsheets of two strings dangling from $u = \infty$ down to
$u=u_{\min}$ at constant $y=\pm L/2$. This contribution is
evaluated by choosing the parametrization $(\tau,\s)=(x^-,u)$,
noting that $\partial_u y=0$. The result is
\be
\label{3-10} \tilde{S}_0 ={\sqrt{f_0} L^-  L \ov 2  \pi } \g \ ,
\ee
and, unlike the cases considered in
\cite{wilsonloop,wilsonloop2,BS} for the heavy quark-antiquark
potential, it is finite. The ``regularized'' action is then $S =
{\rm i} \tilde{S}$ with
\ba
\label{3-11} \tilde{S} & = & \tilde{S}_1 - \tilde{S}_0 =
{\sqrt{f_0} L^-  L \ov 2 \pi } \left(\sqrt{1 + \g^2}-\g\right)
\nonumber\\
& = & {\sqrt{f_0} L^{\! -} L\ov 4\pi} \g^{-1} + {\cal O}(\g^{-3}) = {f_0 L^{\! -} L^2\ov 8 \pi I[g]} + \dots\ ,
\ea
where in the last step we have used \eqn{3-8} and expanded for
small separation distances $L$ or, equivalently, for large $\g$.
Then, the jet quenching parameter can be read off from Eq.
\eqn{1-1} to be
\be
\label{3-13} \hat{q} = {f_0 \ov \pi I[g]}\ .
\ee

\subsection{The case of zero R-charge density}

In this case $r_0=0$ and the various functions appearing in our general expressions become
\be
\label{3-151}
f_0 = {\m^4\ov 2 R^4} \ ,\qq g ={\m^4\ov 2}  {1 \ov u^4 - \m^4 }\ ,
\ee
whereas $h={\m^4\ov 2 u^2}$ mentioned above as being a general feature of our metrics.
Hence,
\be
\label{3-161}
I[g]= {\m^2\ov \sqrt{2}} \int_{\m}^\infty {du \ov \sqrt{u^4 - \m^4} }
= {\m \ov 2} \elK (1/\sqrt{2}) % \simeq 0.93 \m
\ .%= {\m\ov 2} \sqrt{2 \pi} \G({5\ov 4})/ \G({3 \ov 4} )\ .
\ee
For vanishing $r_0$ we have that $\m = \pi R^2 T$ and therefore
\ba
\hat q_0 = {\pi^2 R^2 T^3\ov  \elK(1/\sqrt{2})}
= {\pi^{3/2} \G(3/4)\ov \sqrt{2} \G(5/4)} R^2 T^3 \ .
\label{r0o}
\ea
This is the same as the result derived in \cite{liu}.

\subsection{Two equal nonzero angular momenta}

For the case of two equal nonzero angular momenta, the various functions appearing in
our general expressions are
\be
\label{3-15}
f_0 = {\m^4\ov 2 R^4} \ ,\qq g ={\m^4\ov 2}  {u^2 \ov (u^4 - \m^4)(u^2 - r_0^2)}\ ,
\ee
whereas $h={\m^4\ov 2 u^2}$ as above.
Hence,
\be
\label{3-16}
I[g] = {\m^2\ov \sqrt{2}} \int_{\m}^\infty {du \ u \ov \sqrt{(u^4 - \m^4)(u^2 - r_0^2)} }
= {\m \ov 2} \elK (k)\ ,
\ee
with the modulus $k$ of the elliptic integral being given by
\be
\label{3-17}
k^2 = {1 \ov 2} \left( 1 + {r_0^2 \ov \m^2} \right) = \ha {1 + 2 F^2 \ov 1 + F^2}\ ,
%\qq {1\ov \sqrt{2}} \leqslant k \leqslant 1\ ,
\ee
where we have used \eqn{3-20} to pass from the supergravity to
the gauge-theory parameters and $F$ is understood as $F(\xi)$
in the CE and as $F(\hat \xi)$ in the GCE.
Therefore, the jet quenching parameter is given by
\be
\label{3-22}
\hat{q} ={\m^3 \ov \pi R^4 \elK(k)} =  {\pi^2 R^2 T^3 \ov \elK (k)} \left(1 + F \right)^{3/2}
= {\pi^2 R^2 T^3 \ov \elK (k)}{1\ov (2 k'^2)^{3/2}}\ ,
\ee
where all three different expressions are equivalent and $k'=\sqrt{1-k^2}$
is the complementary modulus.

\vskip 0 cm
\begin{figure}[t]
\begin{center}
\includegraphics [height=7cm]{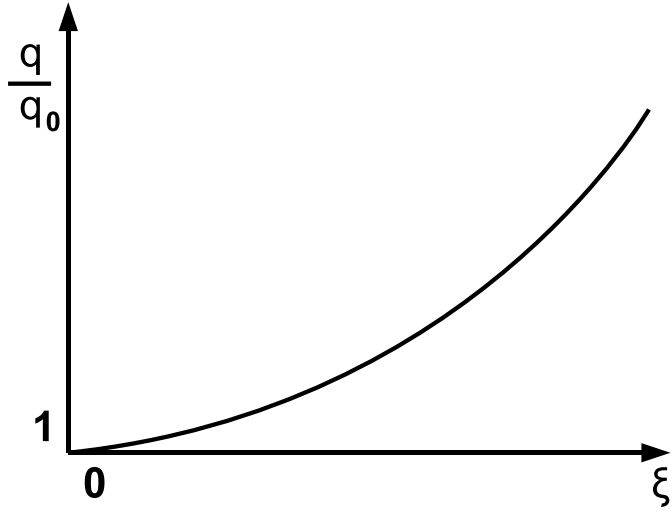}
\end{center}
\vskip -.5 cm \caption{The $\hat{q} / \hat{q}_0$ ratio for the
case of two equal nonzero angular parameters plotted as a function
of the dimensionless parameter
$0\leqslant \xi<\infty $ in \eqn{3-18} appropriate for the CE.
Some indicative values are:
$(\xi, {\hat q\ov \hat q_0}) = (\ha, 1.04), (1, 1.15)$ and $(3,1.76)$.
% = {6 \sqrt{3} \ov  \pi N^2} {J \ov  T^3}$.
The corresponding plot for the GCE in terms of the parameter $0\leqslant \hat
\xi\leqslant 1$ in \eqn{fhi2} is similar in shape with indicative values
$(\hat \xi, {\hat q\ov \hat q_0}) = (\ha, 1.33)$ and $(1, 2.43)$.
}
\label{fig-ratiosphere}
\end{figure}

\no
We would like to compare this result with \eqn{r0o} in the absence of R-charges.
The comparison should be performed at the same temperature and then viewed as a function
of the parameters $\xi$ and $\hat\xi$ for the cases of the CE and the GCE, respectively.
Keeping this in mind, the ratio with that in the non-rotating case is
\be
\label{3-23}
{\hat{q} \ov \hat{q}_0} = {\elK(1 / \sqrt{2}) \ov \elK (k)} \left(1 + F\right)^{3/2}
= {\elK(1 / \sqrt{2}) \ov \elK (k)}{1\ov (2 k'^2)^{3/2}}\ .
\ee
This is a monotonically increasing function of $\xi$ or $\hat\xi$, as is also
demonstrated in Figure 1. For small deviations from unity we have the expansions
\ba
{\rm CE}:\qq {\hat q\ov \hat q_0} & = &1 + 0.188\ \xi^2 - 0.052\ \xi^4
+ 0.026\ \xi^6 + {\cal O}(\xi^8)\ ,
\nonumber\\
{\rm GCE}:\qq {\hat q\ov \hat q_0} & = & 1 + 1.27\ \hat \xi^2+ 0.188\ \hat\xi^4
- 0.038\ \hat\xi^6 + {\cal O}(\hat\xi^8)\ .
\ea

\subsection{One nonzero angular momentum}

For the case of one nonzero angular momentum, we have
\be
\label{3-152}
f_0 = {\m^4\ov 2 R^4} \ ,\qq g ={\m^4\ov 2}  {1\ov u^4 +r_0^2 u^2 -\m^4}
= {\m^4\ov 2}{1\ov (u^2-u_H^2)(u^2+u_+^2)}\ ,
\ee
where
\be
\label{3-25}
u_+^2 = {1 \ov 2} \left( r_0^2 + \sqrt{r_0^4 + 4 \m^4} \right)\ ,
\ee
while $h={\m^4\ov 2 u^2}$ as before.
Then
\be
\label{3-26}
I[g] ={\m^2\ov \sqrt{2}} \int_{u_H}^\infty {du \ov \sqrt{( u^2 - u_H^2 ) ( u^2 + u_+^2)} }
= {\m^2\ov \sqrt{2}} { \elK (k) \ov (r_0^4 + 4 \m^4)^{1/4}}\ ,
\ee
with the modulus $k$ being
\be
\label{3-27}
k^2 = {1 \ov 2} \left( 1 + {r_0^2 \ov \sqrt{r_0^4 + 4 \m^4}} \right)  = {F\ov 2} \ ,
\ee
where we have used \eqn{3-20}.
Therefore, the jet quenching parameter is given by
\be
\label{3-22d}
\hat{q} ={\m^2 (r_0^4+ 4 \m^4)^{1/4} \ov \sqrt{2} \pi R^4 \elK(k)} =  {\pi^2 R^2 T^3 \ov \elK (k)}
F^2 (2-F)^{1/2}
= {\pi^2 R^2 T^3 \ov \elK (k)} (2 k^2)^2 (2 k'^2)^{1/2}\ .
\ee

\no
Again we compare this to the zero R-charge result \eqn{r0o} at fixed common temperature.
We find that the ratio is
\be
\label{3-23d}
{\hat{q} \ov \hat{q}_0} = {\elK(1 / \sqrt{2}) \ov \elK (k)} F^2 (2-F)^{1/2}
= {\elK(1 / \sqrt{2}) \ov \elK (k)} (2 k^2)^2 (2 k'^2)^{1/2}\ .
\ee
As a function of the dimensionless parameter $0\leqslant \xi\leqslant  \xi_{\rm max}$ it is initially
increasing from unity,
then it reaches a maximum at some $\xi_0$, after which it decreases to reach the final value
at $\xi = \xi_{\rm max}$. A similar shape is obtained
also for the plot as a function of $\hat \xi$. These are depicted in Figure 2.
For small deviations from unity, we have the expansions
\ba
{\rm CE:} \qq {\hat q\ov \hat q_0} & = & 1 + 1.27\ \xi^2 - 4.36\ \xi^4 + 22.65\ \xi^6
+ {\cal O}(\xi^8) \ ,
\nonumber\\
{\rm GCE:} \qq {\hat q\ov \hat q_0} & = & 1 + 1.27\ \hat\xi^2 + 0.731\ \hat\xi^4
+ 0.528\ \hat\xi^6 + {\cal O}(\hat\xi^8)\ .
\ea
\vskip 0 cm
\begin{figure}[t]
\begin{center}
\includegraphics[height= 7 cm,angle= 0]{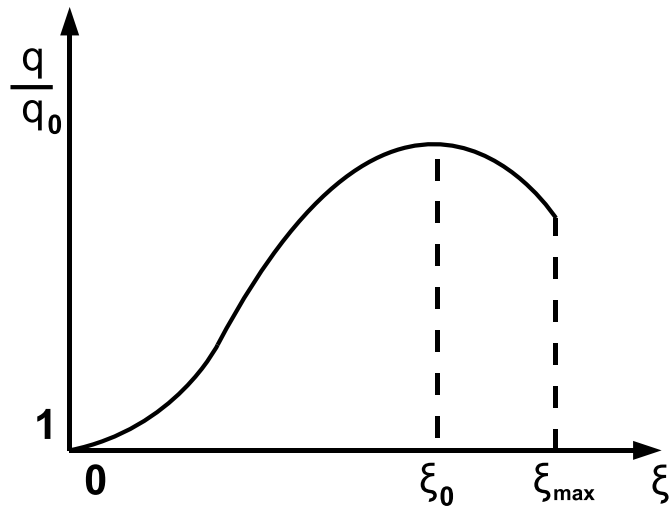}
%\!\!\!\!\!\!\!\!\!\!\!\!\!\!\!\!\!\!\!\!\!\!\!\!\!\!\!\!
%\includegraphics[height= 12 cm,angle= 0]{fig2OLD2.eps}
\end{center}
\vskip -.5 cm
\caption{
The $\hat{q} / \hat{q}_0$ ratio for the case of one nonzero
angular parameter and the CE plotted as a function of the dimensionless parameter
$0\leqslant \xi\leqslant \xi_{\rm max} $ in \eqn{xidef}.
The maximum and final values occurring at
$\xi_0 = 1.09$ and $\xi_{\rm max}=1.33$ are
${\hat q\ov \hat q_0}= 1.37$ and $1.19$, respectively.
The corresponding plot for the GCE
in terms of the parameter $0\leqslant \hat \xi\leqslant \ha$ in \eqn{ol1}
is similar in shape.
The maximum and final values occurring at
$\hat \xi_0 = 0.499$ and $\hat \xi_{\rm max}=\ha$, are
${\hat q\ov \hat q_0}= 1.369$ and $1.368$, respectively.
Note the closeness of the maximum and final values in the GCE case.}
\label{fig-ratiodisc}
\end{figure}

\no As mentioned in the introduction, an alternative, but not
apparently equivalent, description of the problem of energy loss
in the plasma is provided by the drag force felt by a particle
passing through the medium. In this latter context, the authors of
\cite{caceres}, utilizing the same metric \eqn{discc}, showed that
the drag force is an increasing function of the R-charge for small
values of a parameter similar to $\xi$. In this perturbative
regime, this is in qualitative agreement with our results. On the
other hand, we have shown that the ratio $\hat{q} / \hat{q}_0$
reaches a maximum at a finite value of $\xi$ (or $\hat{\xi}$)
lower than its maximal value. In \cite{herzog}, a computation of
the drag force was performed using a five-dimensional black hole
solution \cite{sabra} arising from the dimensionally reduced
spinning D3-brane solutions that we have been using. Indeed, in
that case the drag force exhibits a behavior (see fig. 2 of
\cite{herzog}) sensitive to the number of equal nonzero angular
momenta, which is qualitatively in agreement with expectations
based on our computation of the jet quenching parameter.

\vskip .1in \noindent {\bf Note added.} While this paper was being
typewritten, we received \cite{Caceres:2006as} and
\cite{Lin:2006au} whose results partially overlap with those of
section 3.3.

\vskip .1in

\centerline{ \bf Acknowledgments}

\no
We would like to thank D. Zoakos for helpful discussions.
K.~S. acknowledges support provided through the European
Community's program ``Constituents, Fundamental Forces and
Symmetries of the Universe'' with contract MRTN-CT-2004-005104,
the INTAS contract 03-51-6346 ``Strings, branes and higher-spin gauge fields'',
the Greek Ministry of Education programs $\rm \P Y\Th A\G OPA\S$ with contract 89194
and the program $\rm E\Pi A N$ with code-number B.545.


\begin{thebibliography}{99}

\renewcommand{\baselinestretch}{1}


\bibitem{qgpfluid}
  E.~Shuryak,
  %``Why does the quark gluon plasma at RHIC behave as a nearly ideal fluid?,''
  Prog.\ Part.\ Nucl.\ Phys.\  {\bf 53} (2004) 273,
  {\tt hep-ph/0312227} and
  %%CITATION = HEP-PH 0312227;%%
  %``What RHIC experiments and theory tell us about properties of  quark-gluon
  %plasma?,''
  Nucl.\ Phys.\ {\bf A750} (2005) 64,
  {\tt hep-ph/0405066}. \hfill\break
  %%CITATION = HEP-PH 0405066;%%
  M.J.~Tannenbaum,
  {\it Recent results in relativistic heavy ion collisions: from ``a
  new state of matter'' to ``a perfect fluid''},
  {\tt nucl-ex/0603003}.
  %%CITATION = NUCL-EX 0603003;%%

\bibitem{qgpfluid2}
  D.~Teaney, J.~Lauret and E.~Shuryak,
  {\it A hydrodynamic description of heavy ion collisions at the SPS and RHIC},
  {\tt nucl-th/0110037}. \hfill\break
  %%CITATION = NUCL-TH 0110037;%%
  P.F.~Kolb and U.W.~Heinz,
  {\it Hydrodynamic description of ultrarelativistic heavy-ion collisions},
  {\tt nucl-th/0305084}.
  %%CITATION = NUCL-TH 0305084;%%

\bibitem{adscft}
  J.M.~Maldacena,
  %{\em The large N limit of superconformal field theories and supergravity},
  Adv.\ Theor.\ Math.\ Phys.\  {\bf 2} (1998) 231, Int.\ J.\ Theor.\ Phys.\  {\bf 38} (1999) 1113,
  {\tt hep-th/9711200}. \hfill\break
  %%CITATION = HEP-TH 9711200;%%
  S.S.~Gubser, I.R.~Klebanov and A.M.~Polyakov,
  %{\em Gauge theory correlators from non-critical string theory},
  Phys.\ Lett.\ {\bf B428} (1998) 105,
  {\tt hep-th/9802109}. \hfill\break
  %%CITATION = HEP-TH 9802109;%%
  E.~Witten,
  %{\em Anti-de Sitter space and holography},
  Adv.\ Theor.\ Math.\ Phys.\ {\bf 2} (1998) 253,
  {\tt hep-th/9802150}.
  %%CITATION = HEP-TH 9802150;%%

\bibitem{starinets}
  G.~Policastro, D.T.~Son and A.O.~Starinets,
  %``The shear viscosity of strongly coupled N = 4 supersymmetric Yang-Mills
  %plasma,''
  Phys.\ Rev.\ Lett.\  {\bf 87} (2001) 081601,
  {\tt hep-th/0104066},
  %%CITATION = HEP-TH 0104066;%%
  %``From AdS/CFT correspondence to hydrodynamics,''
  JHEP {\bf 0209} (2002) 043,
  {\tt hep-th/0205052} and
  %%CITATION = HEP-TH 0205052;%%
  %``From AdS/CFT correspondence to hydrodynamics. II: Sound waves,''
  JHEP {\bf 0212} (2002) 054,
  {\tt hep-th/0210220}. \hfill\break
  %%CITATION = HEP-TH 0210220;%%
  A.~Buchel, J.T.~Liu and A.O.~Starinets,
  %``Coupling constant dependence of the shear viscosity in N = 4
  %supersymmetric Yang-Mills theory,''
  Nucl.\ Phys.\ {\bf B707} (2005) 56,\hfill\break
  {\tt hep-th/0406264}. \hfill\break
  %%CITATION = HEP-TH 0406264;%%
  D.T.~Son and A.O.~Starinets,
  %``Hydrodynamics of R-charged black holes,''
  JHEP {\bf 0603} (2006) 052,
  {\tt hep-th/0601157}.
  %%CITATION = HEP-TH 0601157;%%

\bibitem{svuniversality}
  P.~Kovtun, D.T.~Son and A.O.~Starinets,
  %``Holography and hydrodynamics: Diffusion on stretched horizons,''
  JHEP {\bf 0310} (2003) 064,
  {\tt hep-th/0309213},
  %%CITATION = HEP-TH 0309213;%%
  %``Viscosity in strongly interacting quantum field theories from black hole
  %physics,''
  and Phys.\ Rev.\ Lett.\  {\bf 94} (2005) 111601,
  {\tt hep-th/0405231}. \hfill\break
  %%CITATION = HEP-TH 0405231;%%
  A.~Buchel and J.T.~Liu,
  %``Universality of the shear viscosity in supergravity,''
  Phys.\ Rev.\ Lett.\  {\bf 93} (2004) 090602,
  {\tt hep-th/0311175}. \hfill\break
  %%CITATION = HEP-TH 0311175;%%
  A.~Buchel,
  %``On universality of stress-energy tensor correlation functions in
  %supergravity,''
  Phys.\ Lett.\ {\bf B609} (2005) 392,
  {\tt hep-th/0408095}.
  %%CITATION = HEP-TH 0408095;%%

\bibitem{viscosity-rcharge}
  J.~Mas,
  %``Shear viscosity from R-charged AdS black holes,''
  JHEP {\bf 0603} (2006) 016,
  {\tt hep-th/0601144}. \hfill\break
  %%CITATION = HEP-TH 0601144;%%
  K.~Maeda, M.~Natsuume and T.~Okamura,
  %``Viscosity of gauge theory plasma with a chemical potential from AdS/CFT,''
  Phys.\ Rev.\  {\bf D73} (2006) 066013,
  {\tt hep-th/0602010}.
  %%CITATION = HEP-TH 0602010;%%

\bibitem{jqphen}
  J.D.~Bjorken,
  {\it Energy Loss Of Energetic Partons In Quark - Gluon Plasma: Possible
  Extinction Of High P(T) Jets In Hadron - Hadron Collisions},
  FERMILAB-PUB-82-059-THY. \hfill\break
  M.~Gyulassy and M.~Pl\"umer,
  %``Jet Quenching In Dense Matter,''
  Phys.\ Lett.\ {\bf B243} (1990) 432. \hfill\break
  %%CITATION = PHLTA,B243,432;%%
  M.H.~Thoma and M.~Gyulassy,
  %``Quark Damping And Energy Loss In The High Temperature QCD,''
  Nucl.\ Phys.\ {\bf B351} (1991) 491. \hfill\break
  %%CITATION = NUPHA,B351,491;%%
  R.~Baier, D.~Schiff and B.G.~Zakharov,
  %``Energy loss in perturbative QCD,''
  Ann.\ Rev.\ Nucl.\ Part.\ Sci.\  {\bf 50} (2000) 37,
  {\tt hep-ph/0002198}. \hfill\break
  %%CITATION = HEP-PH 0002198;%%
  B. M\"uller,
  %``Phenomenology of jet quenching in heavy ion collisions,''
  Phys.\ Rev.\ {\bf C67} (2003) 061901,
  {\tt nucl-th/0208038}. \hfill\break
  %%CITATION = NUCL-TH 0208038;%%
  M.~Gyulassy, I.~Vitev, X.N.~Wang and B.W.~Zhang,
  {\it Jet quenching and radiative energy loss in dense nuclear matter},
  {\tt nucl-th/0302077}.
  %%CITATION = NUCL-TH 0302077;%%

\bibitem{kovner}
  A.~Kovner and U.A.~Wiedemann,
  %``Eikonal evolution and gluon radiation,''
  Phys. Rev. {\bf D64} (2001) 114002,
  {\tt hep-ph/0106240}.
  %%CITATION = HEP-PH 0106240;%%

\bibitem{wiedemann}
  U.A.~Wiedemann,
  %``Gluon radiation off hard quarks in a nuclear environment: Opacity
  %expansion,''
  Nucl.\ Phys.\ {\bf B588} (2000) 303,
  {\tt hep-ph/0005129}. \hfill\break
  %%CITATION = HEP-PH 0005129;%%
  A.~Kovner and U.A.~Wiedemann,
  {\it Gluon radiation and parton energy loss}, \hfill\break
  {\tt hep-ph/0304151}.
  %%CITATION = HEP-PH 0304151;%%

\bibitem{baier}
  R.~Baier, Y.L.~Dokshitzer, A.H.~Mueller, S.~Peign{\'e} and D.~Schiff,
  %``Radiative energy loss and p(T)-broadening of high energy partons in
  %nuclei,''
  Nucl.\ Phys.\ {\bf B484} (1997) 265,
  {\tt hep-ph/9608322}.
  %%CITATION = HEP-PH 9608322;%%

\bibitem{wilsonloop}
  J.M.~Maldacena,
  %{\em Wilson loops in large $N$ field theories},
  Phys. Rev. Lett.  {\bf 80} (1998) 4859,
  {\tt hep-th/9803002}. \hfill\break
  %%CITATION = HEP-TH 9803002;%%
  S.J.~Rey and J.T.~Yee,
  %``Macroscopic strings as heavy quarks in large N gauge theory and  anti-de
  %Sitter supergravity,''
  Eur.\ Phys.\ J. {\bf C22} (2001) 379,
  {\tt hep-th/9803001}.
  %%CITATION = HEP-TH 9803001;%%

\bibitem{wilsonloop2}
  S.J.~Rey, S.~Theisen and J.T.~Yee,
  %``Wilson-Polyakov loop at finite temperature in large N gauge theory and
  %anti-de Sitter supergravity,''
  Nucl.\ Phys.\ {\bf B527} (1998) 171,
  {\tt hep-th/9803135}. \hfill\break
  %%CITATION = HEP-TH 9803135;%%
  A.~Brandhuber, N.~Itzhaki, J.~Sonnenschein and S.~Yankielowicz,
  %``Wilson loops in the large N limit at finite temperature,''
  Phys.\ Lett.\ {\bf B434} (1998) 36,
  {\tt hep-th/9803137}.
  %%CITATION = HEP-TH 9803137;%%

\bibitem{BS} A.~Brandhuber and K.~Sfetsos,
  %{\em Wilson loops from multicentre and rotating branes, mass gaps and phase
  %structure in gauge theories},
  Adv. Theor. Math. Phys.  {\bf 3} (1999) 851,\hfill\break {\tt hep-th/9906201}.
  %%CITATION = HEP-TH 9906201;%%

\bibitem{liu}
  H.~Liu, K.~Rajagopal and U.~A.~Wiedemann,
  %``Calculating the jet quenching parameter from AdS/CFT,''
  Phys.\ Rev.\ Lett.\  {\bf 97} (2006) 182301,
  {\tt hep-ph/0605178}.
  %%CITATION = HEP-PH 0605178;%%

\bibitem{drag1}
  C.P.~Herzog, A.~Karch, P.~Kovtun, C.~Kozcaz and L.G.~Yaffe,
  %``Energy loss of a heavy quark moving through N = 4 supersymmetric
  %Yang-Mills plasma,''
  JHEP {\bf 0607} (2006) 013,
  {\tt hep-th/0605158}.
  %%CITATION = HEP-TH 0605158;%%

\bibitem{drag2}
  %%CITATION = HEP-TH 0605158;%%
  S.S.~Gubser,
  {\it Drag force in AdS/CFT},
  {\tt hep-th/0605182}.
  %%CITATION = HEP-TH 0605182;%%

\bibitem{diffusion}
  J.~Casalderrey-Solana and D.~Teaney,
  %``Heavy quark diffusion in strongly coupled N = 4 Yang Mills,''
  Phys.\ Rev.\ {\bf D74} (2006) 085012,
  {\tt hep-ph/0605199}.
  %%CITATION = HEP-PH 0605199;%%

\bibitem{mich}
  J.J.~Friess, S.S.~Gubser and G.~Michalogiorgakis,
  %``Dissipation from a heavy quark moving through N = 4 super-Yang-Mills
  %plasma,''
  JHEP {\bf 0609} (2006) 072,
  {\tt hep-th/0605292}.
  %%CITATION = HEP-TH 0605292;%%

\bibitem{herzog}
  C.P.~Herzog,
  %``Energy loss of heavy quarks from asymptotically AdS geometries,''
  JHEP {\bf 0609} (2006) 032,
  {\tt hep-th/0605191}.
  %%CITATION = HEP-TH 0605191;%%

\bibitem{caceres}
  E.~C{\'a}ceres and A.~Guijosa,
  %``Drag force in charged N = 4 SYM plasma,''
  JHEP {\bf 0611} (2006) 077,
  {\tt hep-th/0605235}.
  %%CITATION = HEP-TH 0605235;%%

\bibitem{guijosa}
  M.~Chernicoff, J.A.~Garcia and A.~Guijosa,
  %``The energy of a moving quark-antiquark pair in an N = 4 SYM plasma,''
  JHEP {\bf 0609} (2006) 068,
  {\tt hep-th/0607089}.
  %%CITATION = HEP-TH 0607089;%%

\bibitem{lrwlong}
  H.~Liu, K.~Rajagopal and U.A.~Wiedemann,
  {\em Wilson loops in heavy ion collisions and their calculation in AdS/CFT},
  {\tt hep-ph/0612168}.
  %%CITATION = HEP-PH 0612168;%%

\bibitem{aev}
  P.C.~Argyres, M.~Edalati and J.F.~V\'azquez-Poritz,
  {\em Spacelike strings and jet quenching from a Wilson loop},
  {\tt hep-th/0612157}.
  %%CITATION = HEP-TH 0612157;%%

\bibitem{lrw-wind}
  H.~Liu, K.~Rajagopal and U.A.~Wiedemann,
  {\it An AdS/CFT calculation of screening in a hot wind},
  {\tt hep-ph/0607062}.
  %%CITATION = HEP-PH 0607062;%%

\bibitem{buchel}
  A.~Buchel,
  %``On jet quenching parameters in strongly coupled non-conformal gauge
  %theories,''
  Phys.\ Rev.\ {\bf D74} (2006) 046006,
  {\tt hep-th/0605178}.
  %%CITATION = HEP-TH 0605178;%%

\bibitem{adscftfiniteT}
  E.~Witten,
  %``Anti-de Sitter space, thermal phase transition, and confinement in  gauge
  %theories,''
  Adv.\ Theor.\ Math.\ Phys.\  {\bf 2} (1998) 505,
  {\tt hep-th/9803131}.
  %%CITATION = HEP-TH 9803131;%%

\bibitem{rotatingbranesmore}
  P.~Kraus, F.~Larsen and S.P.~Trivedi,
  %``The Coulomb branch of gauge theory from rotating branes,''
  JHEP {\bf 9903} (1999) 003,
  {\tt hep-th/9811120}.
  %%CITATION = HEP-TH 9811120;%%

\bibitem{cvetic}
  M.~Cvetic and D.~Youm,
  Nucl. Phys. {\bf B477} (1996) 449,
  {\tt hep-th/9605051}
  %%CITATION = HEP-TH 9605051;%%

\bibitem{rs} J.G.~Russo and K.~Sfetsos,
  %{\em Rotating D3 branes and {QCD} in three dimensions},
  Adv.\ Theor.\ Math.\ Phys.\  {\bf 3} (1999) 131,
  {\tt hep-th/9901056}.
  %%CITATION = HEP-TH 9901056;%%

\bibitem{rotatingbranesthermo1}
  S.S.~Gubser,
  %``Thermodynamics of spinning D3-branes,''
  Nucl. Phys. {\bf B551} (1999) 667,
  {\tt hep-th/9810225}.\hfill\break
  R.G.~Cai and K.S.~Soh,
  %``Critical behavior in the rotating D-branes,''
  Mod. Phys. Lett.  {\bf A14} (1999) 1895, {\tt hep-th/9812121}.
  %%CITATION = HEP-TH 9812121;%%

\bibitem{rotatingbranesthermo2}
  M.~Cvetic and S.S.~Gubser,
  %``Thermodynamic stability and phases of general spinning branes,''
  JHEP {\bf 9907} (1999) 010,
  {\tt hep-th/9903132}.
  %%CITATION = HEP-TH 9903132;%%

\bibitem{rotatingbranesthermo3}
  T.~Harmark and N.A.~Obers,
  %``Thermodynamics of spinning branes and their dual field theories,''
  JHEP {\bf 0001} (2000) 008,
  {\tt hep-th/9910036}.
  %%CITATION = HEP-TH 9910036;%%

\bibitem{sabra}
  K.~Behrndt, M.~Cvetic and W.A.~Sabra,
  Nucl. Phys. {\bf B553} (1999) 317, \hfill\break
  {\tt hep-th/9810227}.
  %%CITATION = HEP-TH 9810227;%%

\bibitem{Caceres:2006as}
  E.~C{\'a}ceres and A.~Guijosa,
  {\it On Drag Forces and Jet Quenching in Strongly Coupled Plasmas},
  {\tt hep-th/0606134}.
  %%CITATION = HEP-TH 0606134;%%

\bibitem{Lin:2006au}
  F.L.~Lin and T.~Matsuo,
  %``Jet quenching parameter in medium with chemical potential from AdS/CFT,''
  Phys.\ Lett.\ {\bf B641} (2006) 45,
  {\tt hep-th/0606136}.
  %%CITATION = HEP-TH 0606136;%%


\end{thebibliography}
\end{document}